# Optical properties of AlGaN nanowires synthesized via ion beam techniques

Santanu Parida,[1,a)] P. Magudapathy,[2] A. K. Sivadasan,[1] Ramanathaswamy Pandian,[1] and Sandip Dhara[1,a)]

[1]*Nanomaterials Characterisation and Sensors Section, Surface and Nanoscience Division, Indira Gandhi Centre for Atomic Research, Homi Bhabha National Institute, Kalpakkam-603102, India*

[2]*Accelerator Materials Science Section, Materials Physics Division, Indira Gandhi Centre for Atomic Research, Kalpakkam-603102, India*

*Abstract*

AlGaN plays a vital role in hetero-structure high electron mobility transistor by employing a two-dimensional electron gas and as electron blocking layer in the multi quantum well light emitting diodes. Nevertheless, the incorporation of Al in GaN for the formation of AlGaN alloy is limited by the diffusion barrier formed by instant nitridation of Al adatoms by reactive atomic N. Incorporation of Al above the miscibility limit, however can be achieved by ion beam technique. The well known ion beam mixing (IBM) technique was carried out with the help of $Ar^+$ irradiation for different fluences. A novel approach was also adopted for the synthesis of AlGaN by the process of post irradiation diffusion (PID) as a comparative study with the IBM technique. The optical investigations of AlGaN nanowires, synthesized via two different methods of ion beam processing are reported. The effect of irradiation fluence and post irradiation annealing temperature on the random alloy formation were studied by the vibrational and photoluminescence (PL) spectroscopic studies. Vibrational studies show one-mode phonon behavior corresponding to longitudinal optical (LO) mode of $A_1$ symmetry ($A_1$(LO)) for the wurtzite phase of AlGaN nanowires in the random alloy model. Maximum Al atomic percentage ~6.3-6.7% was calculated with the help of band bowing formalism from the Raman spectral analysis for samples synthesized in IBM and PID processes. PL studies show the extent of defects present in these samples.

[a)]Authors to whom correspondence should be addressed. Electronic addresses: santanuparida026@gmail.com, dhara@igcar.gov.in.

**I. INTRODUCTION**

Group III nitrides and its ternary alloys with direct band gap find tremendous attention in the scientific community as well as the optoelectronic industry for its applications of light emitting diodes (LED) and laser diodes.[1] Among them, AlGaN based LEDs serve as a tunable source in the region of ultra-violet (UV) to deep UV (3.4-6.2 eV) by varying the Al percentage.[2] The deep UV LEDs find potential application in air and water purification, biomedical treatment, high resolution display and optical data storage.[3-5] Generally, the internal quantum efficiency (IQE) of the optoelectronic devices is very low due to high leakage current in InGaN/GaN based blue LEDs. The IQE can be increased to a significant extent by using AlGaN as an electron blocking layer in the multi quantum well (MQW) LEDs.[6] Moreover, AlGaN plays a vital role in hetero-structure high electron mobility transistor by employing a two-dimensional electron gas induced at the hetero-interface as a highly conducting channel.[7]

The synthesis of AlGaN with the desired percentage of Al, however is difficult due to the following reasons. During the growth process, highly reactive atomic N reacts instantly with Al adatoms and limits the further diffusion of the adatoms to form AlGaN with a higher atomic percentage of Al.[8] Furthermore, incorporation of oxygen is promoted due to the high chemical affinity of Al towards oxygen and hence it suppresses the growth of AlGaN.[9] Especially, in the commercially viable technique of chemical vapour deposition (CVD), it is very difficult to achieve high quality AlGaN with a higher percentage (>10%) of Al, because of the miscibility limit and the requirement of very high temperature (>1100 ºC) for the formation of Ga and Al eutectic alloy in the catalyst assisted vapor-liquid-solid (VLS) process.[10] In this context we may like to mention that achieving Al content >3 at% in AlGaN nanostructure using thermodynamic processes is very difficult pertaining to the high value of defect formation energy in nanostructures while incorporation of Al in the lattice of GaN nanostructures.[11] The above mentioned difficulties can be tackled up to a certain extent by adopting the well known ion beam mixing (IBM) techniques, useful for alloy formation above the miscibility limit and can be employed to thermodynamically immiscible systems, as well.[12] In comparison to the direct implantation, IBM results in a very uniform composition throughout the depth of penetration of the irradiation ion. Defect creation and simultaneous diffusion of elemental species are anticipated to form alloy with higher incorporation percentage than the conventional process. Furthermore, IBM is a versatile technique for multi layer formation in metallic and semiconductor systems.[13]

In this present study, we used the IBM technique for the synthesis of AlGaN nanowires (NWs). A new approach was also adopted for the synthesis of AlGaN by the process of post irradiation diffusion (PID), with a motivation for the enhancement of the Al diffusion into GaN due to the production of native defects by the irradiation of high energy ions. Raman spectroscopic study was performed to investigate the vibrational properties and to estimate the Al percentage in the alloy formation. Variation of Al percentage with the irradiation fluence and post irradiation annealing temperature was estimated in both the IBM and the PID processes. Defect creation and subsequent diffusion of elemental species in the later case, may help in forming alloy with higher incorporation percentage than the conventional one. Photoluminescence (PL) spectroscopic study was carried out for understanding the defects present in AlGaN synthesized via the IBM and the PID processes.

## II. EXPERIMENTAL

### A. Synthesis

AlGaN NWs were synthesized by two processes, namely IBM and PID techniques using GaN NWs grown by atmospheric pressure CVD (APCVD) technique in the catalyst assisted vapor-liquid-solid (VLS) process. Au catalyst was deposited on the crystalline Si(100) substrates using thermal evaporation technique (12A4D, HINDHIVAC, India). The Au coated substrates were annealed at a temperature of 900 °C for 10 min in an inert atmosphere for preparing the Au nanoparticles (NPs). Ga metal (99.999%, Alfa Aesar) as precursor, $NH_3$ (5 N pure) as reactant and $N_2$ (5 N pure) as carrier gases were used for the growth process. The Si(100) substrate with Au NPs was kept upstream of a Ga droplet in a high pure alumina boat (99.95%), placed inside a quartz tube. The temperature of the quartz tube was raised to the optimized growth temperature of 900 °C with a ramp rate of 15 °C per min. The growth process was carried out by using $NH_3$ (10 sccm) diluted with $N_2$ (20 sccm) as carrier gas for a duration of 3 h. In IBM process, $Ar^+$ ion was chosen because of its inert nature and to maximize the nuclear energy loss in Al layer. The required thickness of Al layer was calculated as ~25 nm with the help of stopping and range of ions in matter (SRIM) analysis for the $Ar^+$ irradiation with energy of 25 keV. The distribution of the $Ar^+$ in Al/GaN is available in the supplementary material (Fig. S1). Al layer (~25 nm) was coated on as-grown GaN NWs using the same thermal evaporation technique. Al/GaN was irradiated with $Ar^+$ (25 keV) at fluences of 1E16 and 5E16 ions·$cm^{-2}$. The *in-situ* monitoring of fluence for the ion beam was carried out with the help of a Faraday cup. Two separate irradiated samples were selected for annealing process at 900 and 1000 °C in $N_2$ (5N pure) atmosphere for 5 min. On the other hand, for the PID process, prior to the coating of Al (~25 nm thick), the GaN NWs were irradiated with $Ar^+$ ion keeping the energy and fluences of the $Ar^+$ same as in case of IBM process. Similarly, the samples were also annealed with the above mentioned conditions as a final step.

### B. Characterizations

Morphological analysis was carried out after each step using field emission scanning electron microscopy (FESEM; Zeiss SUPRA 55). The vibrational properties were studied in the backscattering geometry using Raman spectroscopy (inVia, Renishaw, UK) with $Ar^+$ laser excitation of 514.5 nm and 1800 groves·mm$^{-1}$ grating used as a monochromatizer. Resonance Raman and photoluminescence spectra were recorded at room temperature using a UV laser of wavelength 325 nm in the same spectrometer. The spectra were collected with a 40× objective and dispersed through a grating of 2400 grooves·mm$^{-1}$. The thermoelectrically cooled charged coupled device (CCD) based detector was used for all these studies.

## III. RESULTS AND DISCUSSION

### A. The morphological analysis

The typical FESEM image of as-grown GaN NWs shows [Figs. 1(a(i)) and 1(b(i))] smooth and uniform NWs with an average diameter of ~100 nm and length varying from 1-4 μm. Lengths of the NWs were calculated from the plan-view, and tilt-view FESEM images (not shown in figure) taken at different locations of the sample. A statistic was made out by measuring the length of several NWs. The maximum numbers of NWs were about 3 μm long. Al/GaN NWs, synthesized in the IBM and the PID processes at two different fluences of 1E16 and 5E16 ions·cm$^{-2}$, after annealing at 1000 °C are shown in Figure 1(a(ii)), 1(a(iii)) and 1(b(ii)), 1(b(iii)) respectively. The corresponding FESEM images of the samples annealed at 900 °C is available in the supplementary material (Fig. S2).

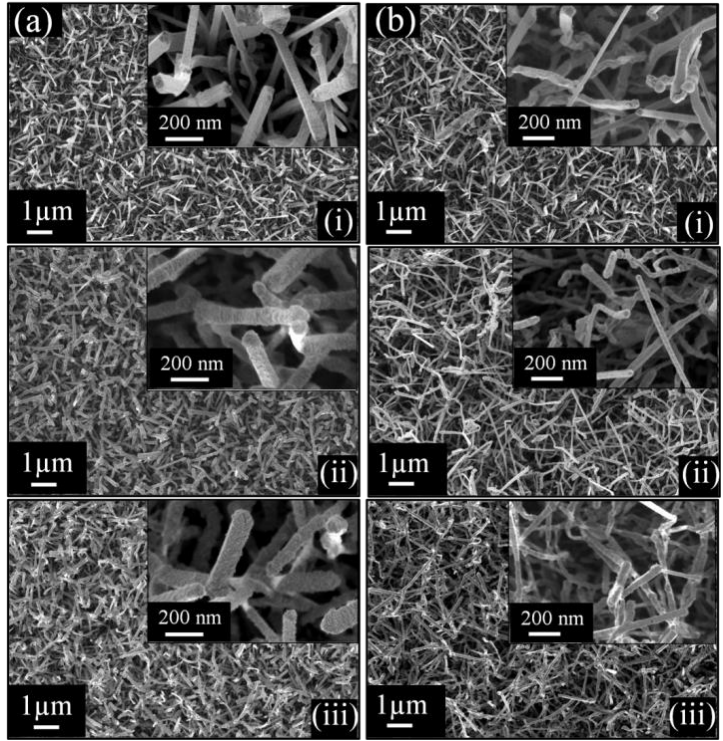

**FIG. 1.** Typical FESEM images for the samples synthesized using IBM and the PID techniques for an Ar$^+$ fluence of (a) 1E16 and (b) 5E16 ions·cm$^{-2}$. (i) as-grown GaN NWs; Al/GaN with post irradiation annealing at 1000 °C in the (ii) IBM and (iii) PID processes. Inserts showing the high magnification images of the respective samples.

The NWs in the samples synthesized via IBM and PID techniques preserve its size, shape and topographic nature after the irradiation with high energy ions with a fluence of 1E16 ions·cm$^{-2}$ and subsequent annealing at 1000 °C, as compared to those for as-grown GaN NWs [Fig. 1(a)]. Similar observations are also made in the case of samples irradiated at a higher fluence of 5E16 ions·cm$^{-2}$ [Fig. 1(b)].

### B. Vibrational analysis

In the case of a wurtzite III-nitride, group theory predicts eight sets of phonon normal modes at the $\Gamma$-point, namely, $2A_1 + 2E_1 + 2B_1 + 2E_2$. Among them, one set of $A_1$ and $E_1$ symmetry modes are acoustic, while the remaining six modes ($A_1 + E_1 + 2B_1 + 2E_2$) are optical in nature. Each one of the $A_1$ and $E_1$ optical modes split into longitudinal optical (LO) and transverse optical (TO) modes due to their polar nature of vibration.[14-16] The typical Raman spectra of as-grown GaN NWs and spectra collected after different steps are shown in the Figure 2 for samples irradiated with a fluence of 1E16 ions·cm$^{-2}$ both in the IBM and the PID processes. In the vibrational spectrum of as-grown GaN NWs, the peaks centered ~567 and ~725 cm$^{-1}$ correspond to the symmetry allowed $E_2$(high) and $A_1$(LO) modes, respectively, for wurzite GaN.[14-16] Another peak centered at ~420 cm$^{-1}$ may correspond to zone boundary phonon modes arising due to the finite crystal size of GaN NWs.[16,17] The peak centered ~520 cm$^{-1}$ is originated from the crystalline Si substrate. In the case of IBM process, after the thin layer deposition of Al, all the Raman modes are found to be quenched [Fig. 2(a)] due the screening of excitation laser by Al metal layer. As a first step in the PID process, however, all the modes are also found to disappear [Fig. 2(b)] after the irradiation with Ar$^+$ because of the production of large number of point defects leading to the distortion of lattice structure. After the annealing at 900 °C in N$_2$ atmosphere, the Raman modes are found to reappear [Figs. 2(a) and 2(b)] due to possible partial removal of defects in the annealing process. In the case of 1000 °C annealed samples, the peaks could revive to their original line shape with significant peak intensity [Figs. 2(a) and 2(b)], as compared to those of the 900 °C annealed samples. However, it could not revive them completely as compared to that of the as–grown sample. The reason may be because of the fact that annealing temperature is not sufficient enough to remove all the point defects created at the time of irradiation. Generally, two third of the material melting point temperature (MPT) is required to remove almost all the point defects created in the irradiation process which is ~1650 °C for GaN (MPT ~ 2500 °C).[18] Moreover, high temperature annealing can also create surface defects, which is significant in nanostructures.[18,19] Therefore, further annealing at higher temperatures were not carried out in order to avoid the creation of surface defects. In the case of post irradiation annealing (900 and 1000 °C) of the sample synthesized in the IBM process, the $E_2$(high) mode exactly reappeared at the same position (~567 cm$^{-1}$), as compared that of the of as-grown sample [Fig. 2(a)]. Similar observations were also found in the case of the sample prepared in the PID process [Fig. 2(b)]. The Raman spectral analysis of the samples synthesized via IBM and PID process with a fluence of 5E16 ions·cm$^{-2}$ also showed the similar characteristics as shown in the supplementary material (Fig. S3).

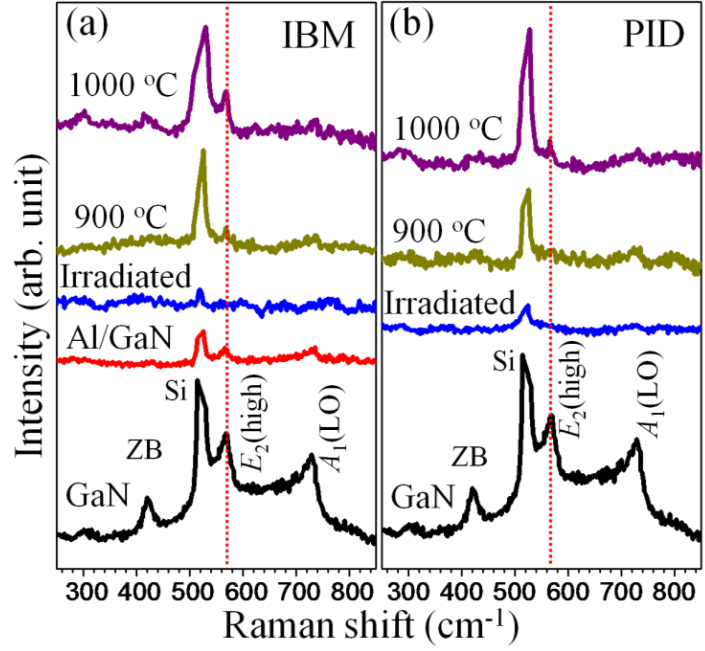

**FIG. 2.** Typical Raman spectra of as-grown GaN and Al/GaN NWs in different steps of the (a) IBM and (b) PID processes irradiated with a fluence of 1E16 ions·cm$^{-2}$. Vertical dashed line is a guide to eye for the peak position of $E_2$(high) mode.

Since, the random alloy formation is expected for the AlGaN;[15] hence $E_2$(high) mode corresponding to AlN is also likely be observed along with the $E_2$(high) mode of GaN. The absence of $E_2$(high) mode corresponding to AlN may be because of the nominal incorporation of Al percentage in GaN, which is not sufficient enough to be probed in the usual Raman spectroscopy, as change in the polarizibility is also very minimum in the AlN phase.[14] Whereas, the $A_1$(LO) mode could not retrieve back the sufficient peak intensity in the Raman spectra even after annealing at 1000 °C. In order to understand the complete behavior of the $A_1$(LO) mode, resonance Raman spectroscopy (RRS) was carried out with the help of 325 nm (3.815 eV) excitation laser sources with energy higher than the band gap of GaN (3.47 eV at room temperature) invoking Frölich interaction in the presence of strong electron-phonon coupling.[20,21] The typical RRS spectra of the samples irradiated with a fluence of 1E16 ions·cm$^{-2}$ in the IBM and the PID process are shown in Figure 3.

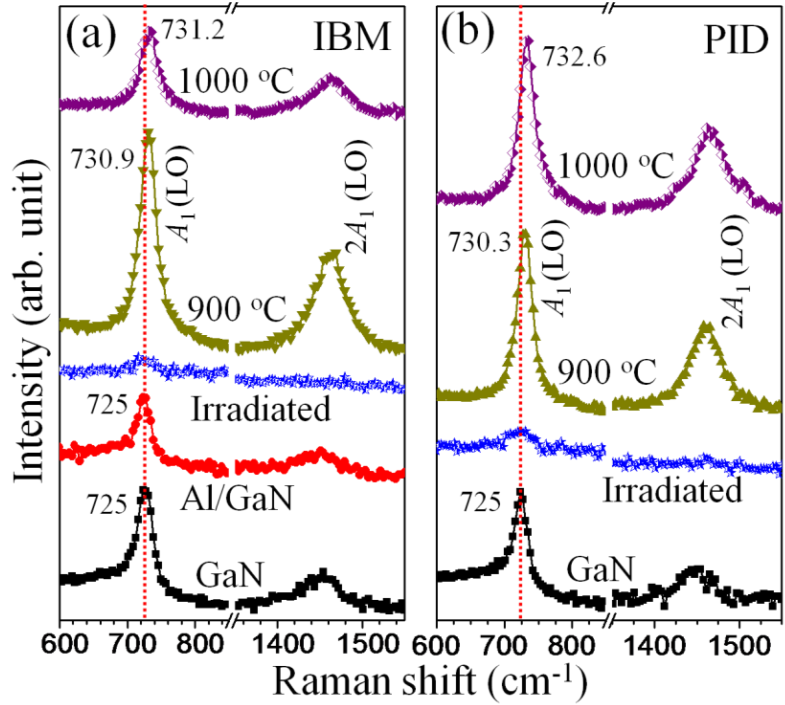

**FIG. 3.** Typical resonance Raman spectra of as-grown GaN and Al/GaN NWs in different steps of the (a) IBM and (b) PID processes irradiated with a fluence of 1E16 ions·cm$^{-2}$. Vertical dashed line is a guide to eye for the blue shift of $A_1$(LO) mode.

Along with the 1$^{st}$ order $A_1$(LO) mode, the 2$^{nd}$ order $2A_1$(LO) mode was also observed in as-grown sample. After the irradiation process, all the $A_1$(LO) modes are found to be diminished. In the IBM process, once the sample was annealed at 900 °C, the modes are found to reappear with a significant blue shift of $A_1$(LO) mode ~5.9 cm$^{-1}$ [Fig. 3(a)] as compared to that of the as-grown GaN NWs. The observed blue shift can be attributed to the one mode behavior of $A_1$(LO) mode in the random alloy model invoked because of the Al incorporation in the GaN.[15] Since, the incorporation of a dopant or elements for alloy formation mainly depends on ion energy and irradiation fluence in the IBM process;[19] the sample annealed at 1000 °C also show the similar blue shift [Fig. 3(a)]. Whereas, in the case of the PID process, the incorporation of Al in GaN depends on ion energy, irradiation fluence and annealing temperature, as well. In contrast to the IBM process, the driving force involved in the PID process for the migration and incorporation of Al in GaN originate from the thermal energy provided by the annealing process. Thus, there is a significant increase in blue shift (~2.3 cm$^{-1}$) observed for $A_1$(LO) mode in the case of 1000 °C annealed sample as compared to that for the 900 °C annealed sample [Fig. 3(b)]. In the case of PID process, the $A_1$(LO) mode shows a higher blue shift as compared to that of the IBM process, indicating possible higher percentage of Al incorporation in 1000 °C annealed sample. Similar analysis were also carried out for the samples synthesized with a fluence of 5E16 ions·cm$^{-2}$ (Fig. S4).

Figure 4 shows the Lorentzian curve fitted $E_2$(high) mode observed in the normal Raman spectra (514.5 nm excitation, Fig. 4(a)) and $A_1$(LO) mode observed in RRS (325 nm excitation, Fig. 4(b)) for as-grown and 1000 °C annealed samples irradiated with a fluence of 1E16 ions·cm$^{-2}$ in the IBM and the PID process.

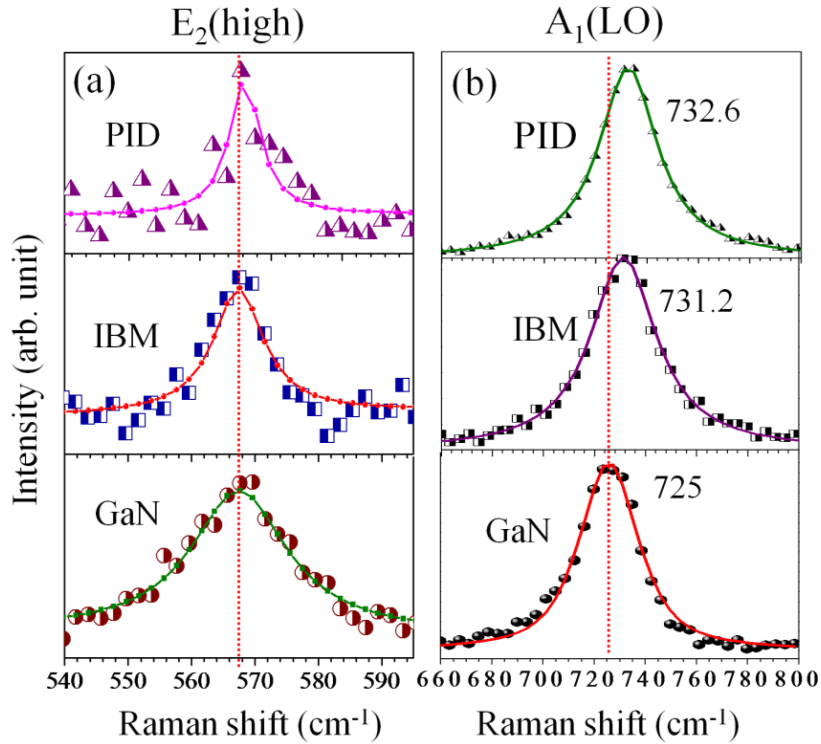

**FIG. 4.** Lorentzian line shape fitted (a) $E_2$(high) and (b) $A_1$(LO) mode of as-grown and 1000 °C annealed samples irradiated with a fluence of 1E16 ions·cm$^{-2}$ in the IBM and the PID processes. Vertical dashed line is a guide to eye for the significant blue shift of $A_1$(LO) mode.

According to band bowing formalism of random alloy model the percentage of Al incorporation is given by the following equation which is analogous to the Vegard's law.[15]

$$A_1(LO)_{AlGaN} = A_1(LO)_{GaN} + [A_1(LO)_{AlN} - A_1(LO)_{GaN}]x - bx(1-x) \quad \ldots\ldots\ldots\ldots\ldots\ldots(1)$$

Where, '$x$' is the Al atomic percentage in AlGaN and '$b$' is the bowing parameter. For lower percentage of Al, the bowing parameter '$b$' can be neglected. Thus, the atomic percentage of Al in the AlGaN random alloy was found to ~3.7 % and ~4.6 % in case of the IBM and the PID processes, respectively for the samples irradiated with a fluence of 1E16 ions·cm$^{-2}$. Similarly, the Al percentage was also estimated as ~6.7 and ~6.3 at% for the samples irradiated with a fluence of 5E16 ions·cm$^{-2}$ in the IBM as well as the PID processes. The incorporation of Al percentage with different fluencies and processes are tabulated (Table I).

TABLE I: Al atomic percentage in AlGaN with different fluences and annealing temperatures for the IBM and the PID processes.

| Ar+ fluence | Al at% in IBM | | Al at% in PID | |
|---|---|---|---|---|
| | 900 °C | 1000 °C | 900 °C | 1000 °C |
| 1E16 ions·cm$^{-2}$ | 3.6 | 3.7 | 3.1 | 4.6 |
| 5E16 ions·cm$^{-2}$ | 6.2 | 6.7 | 5.9 | 6.3 |

The Al incorporation percentage increases with ion fluence in both the processes. In the IBM process, the Al atomic percentage in AlGaN increases ~3 % with increasing the fluence from 1E16 to 5E16 ions·cm$^{-2}$. Whereas, in the case of PID process, increase in atomic percentage of Al with the ion fluence is limited to ~1.7 %. Therefore, it may be concluded that the IBM is more suitable for the random alloy formation as compared to that of with the PID process.

### C. Luminescence Properties

In order to investigate the optical properties, PL spectra for samples irradiated with fluences 1E16 and 5E16 ions·cm$^{-2}$ in both the IBM and the PID processes were also recorded. The typical PL spectra of as-grown and the sample at different conditions are shown in the Figure 5.

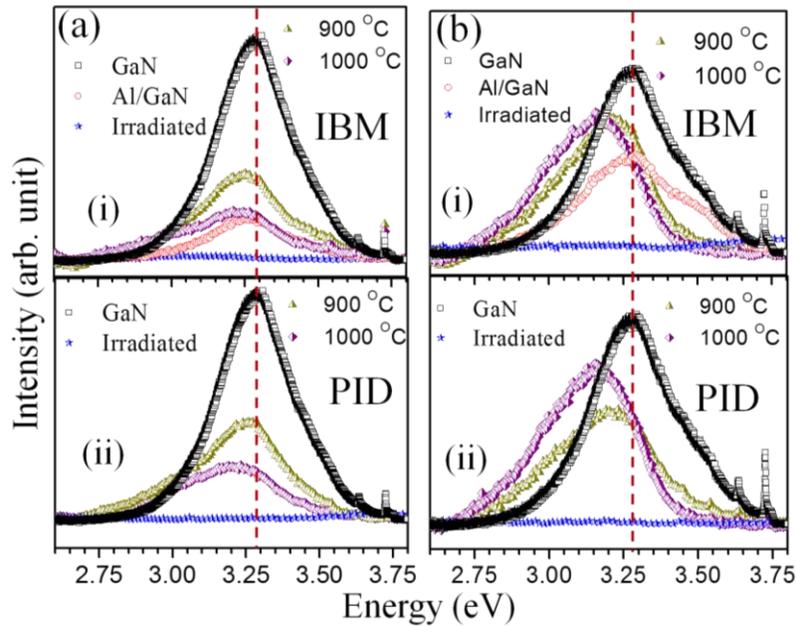

**FIG. 5.** Typical PL spectra of as-grown GaN and Al/GaN NWs for different Ar+ fluences of (a) 1E16 and (b) 5E16 ions·cm$^{-2}$ synthesized with different processes of (i) IBM (ii) PID. Vertical dashed line is a guide to eye for the red shift of DAP peak position.

The as-grown GaN NWs show a peak centered at ~3.28 eV, which is attributed to the recombination of the neutral donor-acceptor pair (DAP), originating due to a transition of electrons from the possible presence of a shallow donor state of nitrogen vacancy ($V_N$) to a deep acceptor state of Ga vacancy ($V_{Ga}$).[22,23] The quenching of PL emission in the irradiated sample (Fig. 5) is attributed to the trapping of radiative charge carriers by large number of point defects produced by the high energy ion irradiation. The DAP emission peak is found to evolve with post irradiation annealing at 900 and 1000 °C in both IBM and PID processes. However, for sample irradiated with a fluence of 1E16 ions·cm$^{-2}$ in the case of the IBM process, the evolved DAP emission shows [Fig. 5a(i)] a red shift of ~30 and 50 meV for 900 and 1000 °C annealed samples, respectively. Similarly the DAP emission for sample irradiated with a fluence of 1E16 ions·cm$^{-2}$ in case of the sample synthesized in PID process [Fig. 5a(ii)] also shows a red shift of ~30 and 60 meV for 900 and 1000 °C annealing, respectively. The increase in defect density at the time of irradiation leads to the creation of a large number of $V_N$ with a possible generation of a band of defect related energy levels possessing a lower energy state than that of the as-grown GaN. Therefore, the red shift observed for the DAP emission may be attributed to the transition from deep defect levels instead of shallow donor levels.[24] Similar observations were found in the PL studies performed on the as-grown sample and the sample irradiated with higher fluence of 5E16 ions·cm$^{-2}$ [Figs. 5b(i) and 5b((ii)]. Moreover, both the 900 and 1000 °C annealed sample show higher red shift of the DAP emission peak observed for the sample irradiated with a fluence of 5E16 ions·cm$^{-2}$ as compared to that for the samples synthesized at a fluence of 1E16 ions·cm$^{-2}$. While ion irradiation can create new defects in some materials, it simultaneously anneal out pre-existing defects in some others (dynamic annealing) with respect to ion energy and fluence.[25] Therefore the further red shift of DAP emission in the samples irradiated with a higher fluence of 5E16 ions·cm$^{-2}$ can be attributed to the creation of deep defect states by annealing out the existed shallow defect states of GaN. Furthermore, the relative intensity of DAP emission of AlGaN synthesized in both IBM and PID processes with an irradiance fluence of 5E16 ions·cm$^{-2}$ is greater than that of the samples synthesized with a fluence of 1E16 ions·cm$^{-2}$. This can be attributed to the dynamic annealing of the defects at the higher fluence of 5E16 ions·cm$^{-2}$.[26]

**IV. CONCLUSIONS**

In summary, AlGaN nanowires (NWs) were synthesized by the ion beam mixing (IBM) and the post irradiation diffusion (PID) processes using GaN NWs grown in the catalyst assisted chemical vapor deposition technique. Vibrational studies with the help of resonance Raman spectroscopy (RRS) for the post irradiation annealed Al/GaN sample showed the one-mode phonon behavior corresponding to $A_1$(LO) mode of the random alloy. The amount of Al incorporation estimated from the RRS studies in the AlGaN sample increases with ion fluence in both the processes. Highest Al atomic percentage of ~6.3-6.7% is achieved in AlGaN at a fluence of 5E16 ions·cm$^{-2}$ in the ion irradiation process. Moreover, gradual red shift of donor acceptor pair (DAP) emission with increasing irradiation fluence and post irradiation annealing temperature in AlGaN shows the creation of deep level defect states in the Al/GaN matrix as compared to that of GaN.

**SUPPLEMENTRY MATERIALS**

See supplementary materials for the distribution of Ar$^+$ irradiated on Al/GaN (FIG. S1), typical FESEM images for the 900 ºC post annealed samples synthesized using ion beam techniques (FIG. S2), typical Raman spectra of GaN and Al/GaN NWs in different steps of the ion beam processes synthesized at a fluence of 5E16 ions·cm$^{-2}$ (FIG. S3) and typical resonance Raman spectra of GaN and Al/GaN NWs in different steps of the ion beam processes synthesized at a fluence of 5E16 ions·cm$^{-2}$ (FIG. S4).

**ACKNOWLEDGEMENTS**

We thank A. Das, Kishore K. Madapu and A. Patsha of SND, IGCAR, for their valuable suggestions and useful discussions.

Table I: Al atomic percentage in AlGaN with different fluences and annealing temperatures for the IBM and the PID processes.

| Ar$^+$ fluence | Al at% in IBM | | Al at% in PID | |
|---|---|---|---|---|
| | 900 °C | 1000 °C | 900 °C | 1000 °C |
| 1E16 ions·cm$^{-2}$ | 3.6 | 3.7 | 3.1 | 4.6 |
| 5E16 ions·cm$^{-2}$ | 6.2 | 6.7 | 5.9 | 6.3 |

**Figure captions**

**FIG. 1.** Typical FESEM images for the samples synthesized using IBM and the PID techniques for an Ar$^+$ fluence of (a) 1E16 and (b) 5E16 ions·cm$^{-2}$. (i) as-grown GaN NWs; Al/GaN with post irradiation annealing at 1000 °C in the (ii) IBM and (iii) PID processes. Inserts showing the high magnification images of the respective samples.

**FIG. 2.** Typical Raman spectra of as-grown GaN and Al/GaN NWs in different steps of the (a) IBM and (b) PID processes irradiated with a fluence of 1E16 ions·cm$^{-2}$. Vertical dashed line is a guide to eye for the peak position of $E_2$(high) mode.

**FIG. 3.** Typical resonance Raman spectra of as-grown GaN and Al/GaN NWs in different steps of the (a) IBM and (b) PID processes irradiated with a fluence of 1E16 ions·cm$^{-2}$. Vertical dashed line is a guide to eye for the blue shift of $A_1$(LO) mode.

**FIG. 4.** Lorentzian line shape fitted (a) $E_2$(high) and (b) $A_1$(LO) mode of as-grown and 1000 °C annealed samples irradiated with a fluence of 1E16 ions·cm$^{-2}$ in the IBM and the PID processes. Vertical dashed line is a guide to eye for the significant blue shift of $A_1$(LO) mode.

**FIG. 5.** Typical PL spectra of as-grown GaN and Al/GaN NWs for different Ar$^+$ fluences of (a) 1E16 and (b) 5E16 ions·cm$^{-2}$ synthesized with different processes of (i) IBM (ii) PID. Vertical dashed line is a guide to eye for the red shift of DAP peak position.

**Supplementary Materials**

**Optical properties of AlGaN nanowires synthesized via ion beam techniques**

Santanu Parida, P. Magudapathy, A. K. Sivadasan, Ramanathaswamy Pandian, Sandip Dhara

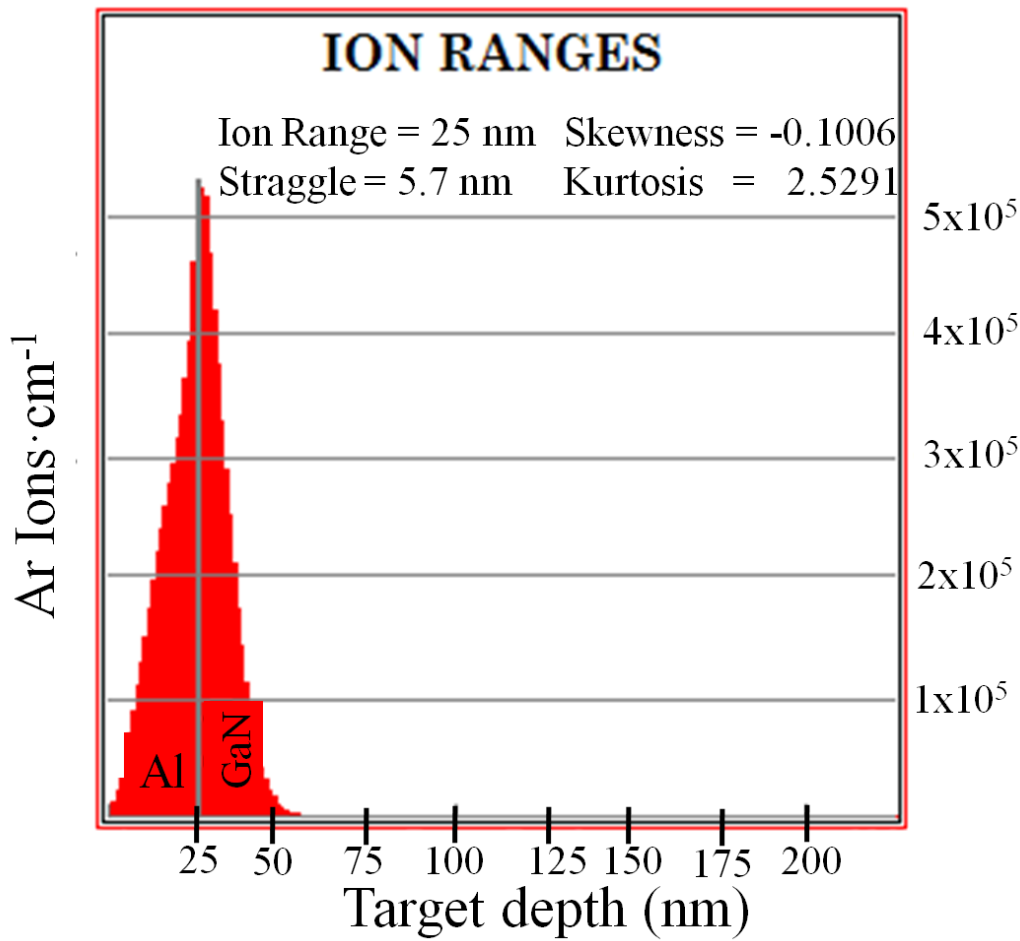

**FIG. S1.** Distribution of 25 keV Ar$^+$ irradiated on Al (25 nm)/GaN.

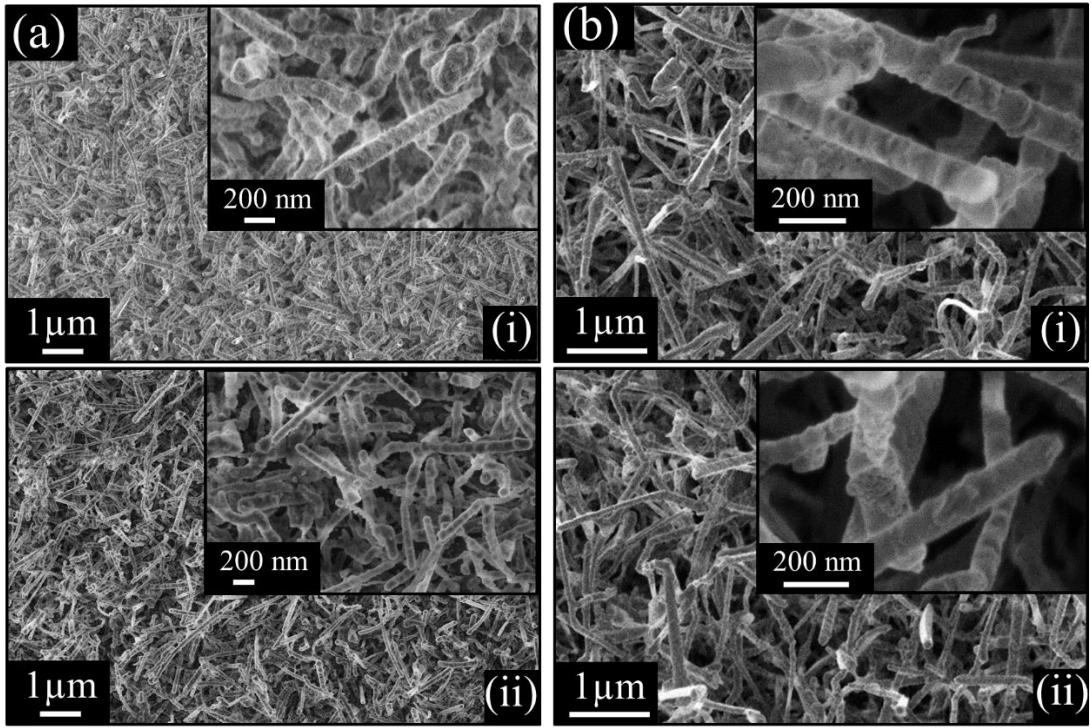

**FIG. S2.** Typical FESEM images for the samples synthesized using IBM and the PID techniques for an Ar$^+$ fluence of (a) 1E16 and (b) 5E16 ions·cm$^{-2}$. (i) Al/GaN with post irradiation annealing at 900 °C in the (i) IBM and (ii) PID processes. Inserts showing the high magnification images of the respective samples.

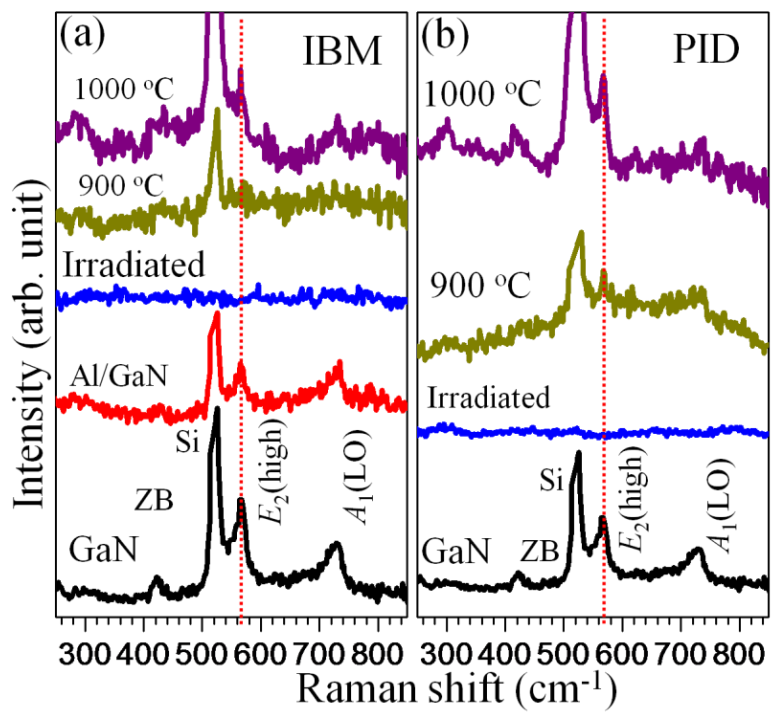

**FIG. S3.** Typical Raman spectra of GaN and Al/GaN NWs in different steps of the (a) IBM and (b) PID processes synthesized at a fluence of 5E16 ions·cm$^{-2}$. Vertical dashed line is a guide to eye for the peak position of the $E_2$(high) mode.

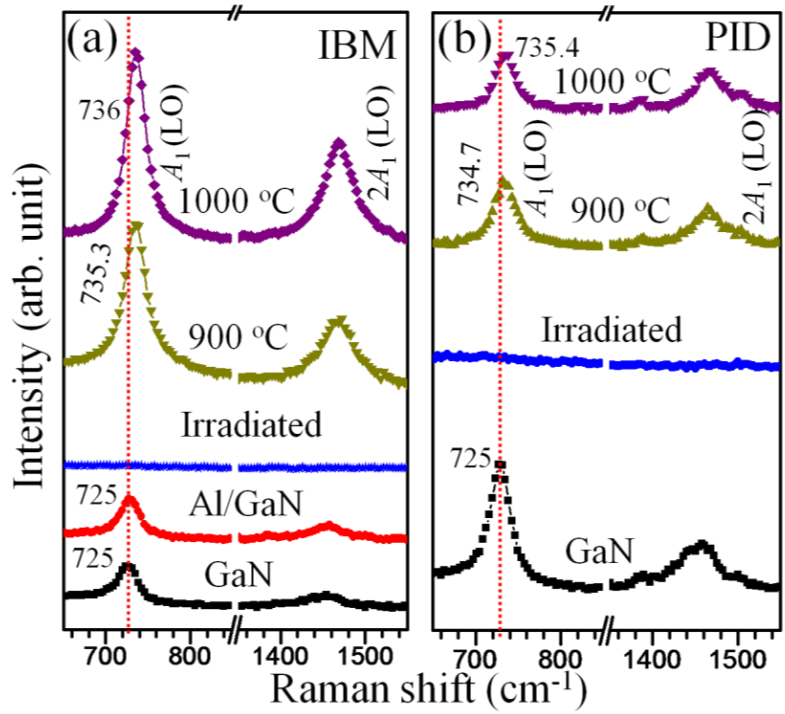

**FIG. S4.** Typical resonance Raman spectra of GaN and Al/GaN NWs in different steps of the (a) IBM and (b) PID processes synthesized at a fluence of 5E16 ions·cm$^{-2}$. Vertical dashed line is a guide to eye for the blue shift of $A_1$(LO) mode.

.